\documentclass[12pt]{article}
\usepackage{amsmath}

\columnwidth\textwidth

\begin{document}

\newcommand{\be}{\begin{equation}}
\newcommand{\ee}{\end{equation}}
\newcommand{\beann}{\begin{eqnarray*}}
\newcommand{\eeann}{\end{eqnarray*}}
\newcommand{\bea}{\begin{eqnarray}}
\newcommand{\eea}{\end{eqnarray}}
\newcommand{\nn}{\nonumber}
\newcommand{\ben}{\begin{enumerate}}
\newcommand{\een}{\end{enumerate}}
\newtheorem{df}{Definition}
\newtheorem{thm}{Theorem}
\newtheorem{lem}{Lemma}
\newtheorem{prop}{Proposition}
\begin{titlepage}

\noindent
\hspace*{11cm} BUTP-00/14 \\
\vspace*{1cm}
\begin{center}
{\LARGE Unitary dynamics \\[0.5cm]
of spherical null gravitating shells} 

\vspace{2cm}

P. H\'{a}j\'{\i}\v{c}ek, \\
Institute for Theoretical Physics, \\
University of Bern, \\
Sidlerstrasse 5, CH-3012 Bern, Switzerland, \\
E-mail: hajicek@itp.unibe.ch, \\
Telephone: (41)(31) 631 86 25, FAX: 631 38 21.
\vspace*{2cm}

\nopagebreak[4]

\begin{abstract}
  The dynamics of a thin spherically symmetric shell of zero-rest-mass matter
  in its own gravitational field is studied. A form of action principle is
  used that enables the reformulation of the dynamics as motion on a fixed
  background manifold. A self-adjoint extension of the Hamiltonian is obtained
  via the group quantization method. Operators of position and of direction of
  motion are constructed. The shell is shown to avoid the singularity, to
  bounce and to re-expand to that asymptotic region from which it contracted;
  the dynamics is, therefore, truly unitary. If a wave packet is sufficiently
  narrow and/or energetic then an essential part of it can be concentrated
  under its Schwarzschild radius near the bounce point but no black hole
  forms. The quantum Schwarzschild horizon is a linear combination of a black
  and white hole apparent horizons rather than an event horizon.
\end{abstract}
\vskip 0.4cm
{\em PACS:} 0460

{\em Keywords:} unitarity in quantum gravity, collapse, black holes, thin
shells 

\end{center}

\end{titlepage}

\section{Introduction}
\label{sec:intro}
According to general relativity, all parts of a massive object definitely
disappear if the object falls through its Schwarzschild radius. The problem to
be tackled in the present paper is whether also a quantum system is, or is
not, irretrievably lost if it falls under its Schwarzschild radius.

We limit ourselves to a sufficiently simple model so that no approximations
are needed and the quantum theory can be constructed without problems. In this
way, an important question about the validity of approximative methods such as
WKB expansion can also be touched. The simplest system that can ever be
invented for these aims seems to be a thin shell with its own gravitational
field made of light-like material, everything spherically symmetric. A
Hamiltonian action principle \cite{L-W-F} for this system has been transformed
to a form suitable for quantization in \cite{H-kief} (foregoing paper); this
will be used as a starting point. Most of the results of the present paper
have already been published in a short review \cite{sard}; here, all
derivations and calculations will be described in sufficient detail, some
new results will be added, and a new interpretation of the results will be
given. 

The plan of the paper is as follows. In Sec.~2, the starting assumptions and
equations are collected. The action for the system from Ref.~\cite{L-W-F} is
written down because we shall need the form of the constraints. The same
action after the transformation to a set of embedding variables and Dirac
observables is then given, the key notion of background manifold is
introduced, and the meaning of the new variables is discussed. A construction
of quantum mechanics including the position and the direction-of-motion
operators is contained in Sec.~3. The so-called group-theoretical quantization
method is used, which is well adapted to the problems such as the limited
ranges of spectra and the construction of a unitary dynamics. The quantum
mechanics is formulated as a dynamics of the shell on the background manifold;
this enables straightforward and unique interpretations. In Sec.~4, motion of
wave packets is investigated. It turns out that no shell reaches the zero
radius if it starts away from it and so the singularity is avoided. The wave
packets contract, bounce and then expand, reaching the asymptotic region from
which they have been sent in, so the dynamics is unitary from the point of
view of one family of observers. Some of the packets can be sufficiently
concentrated near their bouncing point so that an essential part of them comes
under the corresponding Schwarzschild radius, but no event horizon forms.

In Sec.~5, we consider the seemingly contradictory claims that the quantum
shell can cross its Schwarzschild radius and still re-expand. The solution of
the paradox is that if the matter creates a Schwarzschild (apparent) horizon
outside then the horizon can be, even in the classical version of the theory,
of two types: white or black, that is, corresponding to the white or black
hole horizon in the Schwarzschild spacetime. The ``colour'' of the apparent
horizon in a Cauchy surface depends on the direction of motion of the shell:
the horizon is black if the shell is contracting and it is white if the shell
is expanding. The quantum horizon is a linear combination of both because the
motion of the shell is. The quantum horizon is ``grey'', changing from
mostly black to mostly white.

The semi-classical approximation fails blatantly near the bouncing
point of the quantum shell because every classical shell reaches its
Schwarzschild radius, forms a black hole and falls into the singularity. A
cautious discussion of this point is given in Sec.~6. In particular, the
reason is explained why our results do not prevent massive quantum systems
from collapsing to black-hole-like objects.

\section{Canonical formalism}
In this section, we shall summarize the formulae derived in Refs.\ 
\cite{L-W-F} (abbreviated as LWF further on) and \cite{H-kief} that are needed
to start the present paper.

In LWF, the spherically symmetric metric outside the shell is
written in the form
\[
  ds^2 = -N^2d\tau^2 + \Lambda^2(d\rho + N^rd\tau)^2 + R^2d\Omega^2,
\]
and the shell is described by its radial coordinate $\rho = {\mathbf
  r}$. The LWF action reads
\[
  S_0 = \int d\tau\left[{\mathbf p}\dot{\mathbf r} + \int_0^\infty 
  d\rho(P_\Lambda\dot{\Lambda} + P_R\dot{R} - H_0)\right],
\]
and the LWF Hamiltonian is
\[
  H_0 = N{\mathcal H} + N^\rho{\mathcal H}_\rho + N_\infty M_\infty,
\] 
where $N_\infty := \lim_{\rho\rightarrow\infty}N(\rho)$, $M_\infty$ is the ADM
energy, $\mathcal H$ and ${\mathcal H}_\rho$ are the constraints,
\begin{eqnarray}
 {\mathcal H} & = & \frac{\Lambda P_\Lambda^2}{2R} -
 \frac{P_\Lambda P_R}{R} + 
 \frac{RR''}{\Lambda} - \frac{RR'\Lambda'}{\Lambda^2} 
 + \frac{R^{\prime 2}}{2\Lambda} - \frac{\Lambda}{2} + \frac{\eta{\mathbf
 p}}{\Lambda}\delta(\rho - {\mathbf r}),
\label{LWF-H} \\
  {\mathcal H}_\rho & = & P_RR' - P_\Lambda'\Lambda - {\mathbf
 p}\delta(\rho - {\mathbf r});
\label{LWF-Hr}
\end{eqnarray}
the prime denotes the derivative with respect to $\rho$ and the dot that with
respect to $\tau$.

In Ref.\ \cite{H-kief}, the variables $\eta$, $\mathbf r$, $\mathbf p$,
$\Lambda$, $P_\Lambda$, $R$ and $P_R$ have been transformed to the embedding
variables $U(\rho)$ and $V(\rho)$, their canonical conjugates $P_U(\rho)$ and
$P_V(\rho)$, and the shell variables $u$, $v$, $p_u$ and $p_v$. The pair
$\Bigl(U(\rho),V(\rho)\Bigr)$ defines an embedding of the half-axis into the
so-called background manifold $\mathcal M$ that is covered by the coordinates
$U$ and $V$ with the ranges
\[
 \frac{U+V}{2} \in (-\infty,\infty),\quad \frac{-U+V}{2} \in (0,\infty).
\]
The transformation to embedding variables is determined by a gauge condition,
and there has been a definite condition used in Ref.\ \cite{H-kief}, where all
details are given. The background manifold carries then a set of metrics, one
representative for each geometry. The variables $u$ and $v$ are the
coordinates of the shell trajectory in the background manifold:
\[
  U = u(\tau),\quad V = v(\tau).
\]

The full action that results has the form of the so-called {\em Kucha\v{r}
  decomposition}
\begin{equation}
  S = \int d\tau\left(p_u\dot{u} + p_v\dot{v} - np_up_v\right)
  + \int d\tau\int_0^\infty d\rho(P_U\dot{U} + P_V\dot{V} - H),
\label{KD}
\end{equation}
where $H = N^UP_U + N^VP_V$; $N^U(\rho)$ and $N^V(\rho)$ are Lagrange
multipliers. 

The variables $u$, $v$, $p_u$ and $p_v$ span an extended phase space of the
shell. They contain all true degrees of freedom of the system. The phase space
has non-trivial boundaries: 
\begin{equation}
  p_u \leq 0,\quad p_v \leq 0,\quad \frac{-u+v}{2} \geq 0.
\label{boundar}
\end{equation}
The constraint surface of the extended action of the shell consists of two
components: outgoing shells for $p_v = 0$ and in-going shells for $p_u =
0$.

\section{Group quantization}
\label{sec:quant}
To quantize the system defined by the action (\ref{KD}), we apply the
so-called group-theoretical quantization method \cite{isham}. There are three
reasons for this choice. First, the method as modified for the generally
covariant systems by Rovelli \cite{rovel} (see also \cite{honnef} and
\cite{H-I}) is based on the algebra of Dirac observables of the system;
dependent degrees of freedom don't influence the definition of Hilbert space.
Second, the group method has, in fact, been invented to cope with restrictions
such as Eq.\ (\ref{boundar}). Finally, the method automatically leads to
self-adjoint operators representing all observables. 

In particular, a unique self-adjoint extension of the Hamiltonian is obtained
in this way, and this is the reason that the dynamics is unitary. The
uniqueness of the self-adjoint extension of the Hamiltonian is truly a result
of the group quantization in the sense that the Hamiltonian operator itself,
as calculated from the constraint, possesses a one-dimensional family of such
extensions.

To begin with, we have to find a complete system of Dirac observables. Let us
choose the functions $p_u$, $p_v$, $D_u :=up_u$ and $D_v :=vp_v$. Observe that
$u$ alone is constant only along outgoing shell trajectories ($p_u \neq 0$),
and $v$ only along in-going ones ($p_v \neq 0$), but $up_u$ and $vp_v$ are
always constant. The only non vanishing Poisson brackets are
\[
  \{D_u,p_u\} = p_u,\quad \{D_v,p_v\} = p_v.
\]
This Lie algebra generates a group $G_0$ of symplectic transformations of the
phase space that preserve the boundaries $p_u = 0$ and $p_v=0$. $G_0$ is the
Cartesian product of two copies of the two-dimensional affine group $\mathcal A$.

The group $\mathcal A$ generated by $p_u$ and $D_u$ has three
irreducible unitary representations. In the first one, the spectrum of
the operator $\hat{p}_u$ is $[0,\infty)$, in the second, $\hat{p}_u$
is the zero operator, and in the third, the spectrum is $(-\infty,0]$, see
Ref.\ \cite{raczka}. Thus, we must choose the third representation; this can
be described as follows (details are given in Ref.\ \cite{raczka}).

The Hilbert space is constructed from complex functions $\psi_u(p)$
of $p\in [0,\infty)$; the scalar product is defined by
\[ 
  (\psi_u,\phi_u) := \int_0^\infty\ \frac{dp}{p}\ \psi^*_u(p)\phi_u(p),
\]
and the action of the generators $\hat{p}_u$ and $\hat{D}_u$ on smooth
functions is
\[
  (\hat{p}_u\psi_u)(p) = -p\psi_u(p),\quad (\hat{D}_u\psi_u)(p) =
  -ip\frac{d\psi_u(p)}{dp}.
\]
Similarly, the group generated by $p_v$ and $D_v$ is represented on
functions $\psi_v(p)$; the group $G_0$ can, therefore, be represented
on pairs $\Bigl(\psi_u(p),\psi_v(p)\Bigr)$ of functions:
\begin{eqnarray*}
\hat{p}_u\Bigl(\psi_u(p),\psi_v(p)\Bigr) & = & \Bigl(-p\psi_u(p),0\Bigr), \\ 
\hat{p}_v\Bigl(\psi_u(p),\psi_v(p)\Bigr) & = & \Bigl(0,-p\psi_v(p)\Bigr), \\
\hat{D}_u\Bigl(\psi_u(p),\psi_v(p)\Bigr) & = & \Bigl(-ip\frac{d\psi_u(p)}{dp},0\Bigr), \\
\hat{D}_v\Bigl(\psi_u(p),\psi_v(p)\Bigr) & = & \Bigl(0,-ip\frac{d\psi_v(p)}{dp}\Bigr).  
\end{eqnarray*} 
This choice guarantees that the Casimir operator $\hat{p}_u\hat{p}_v$
is the zero operator on this Hilbert space, and so the constraint is
satisfied.

Handling the last inequality (\ref{boundar}) is facilitated by the canonical
transformation:
\begin{alignat}{2}
  t & = (u+v)/2, & \qquad r & = (-u+v)/2,
\label{tr} \\
  p_t & = p_u + p_v, & \qquad p_r & = -p_u + p_v.
\label{ptpr}
\end{alignat}
The constraint function then becomes $p_up_v = (p_t^2 - p_r^2)/4$.

The positivity of $r$ is simply due to its role as the radius of the
shell: it is defined as a square root of a sum of squares of
coordinates with the range ${\mathbf R}^3$. This suggests the
following trick. Let us extend the phase space so that $r \in
(-\infty,+\infty)$ and let us define a symplectic map $I$ on this
extended space by $I(t,r,p_t,p_r) = (t,-r,p_t,-p_r)$. The quotient of
the extended space by $I$ is isomorphic to the original space, and we
adopt it as our phase space.

Clearly, only those functions on the extended space that are invariant
with respect to $I$ will define functions on the quotient. Dirac
observables of this kind are, eg., $p_t$, $p_r^2$, the ``dilation'' $D
:= tp_t + rp_r = up_u + vp_v$ and the square of the ``boost'' $J^2 :=
(tp_r + rp_t)^2 = (-up_u + vp_v)^2$. The ``action'' of the map $I$ on
the functions $p_u$, $p_v$, $D_u$ and $D_v$ is:
\[
  Ip_uI = p_v,\quad ID_uI = D_v,\quad Ip_vI = p_u,\quad ID_vI = D_u.
\]
There are only two choices for $\hat{I}$ that preserve these relations
in the quantum theory:
\[
  \hat{I}\Bigl(\psi_u(p),\psi_v(p)\Bigr) = \Bigl(\pm\psi_v(p),\pm\psi_u(p)\Bigr).
\]
We choose the plus sign; it is easy to see that the other choice leads to an
equivalent theory. Observe that the resulting representation of the group $G
:= G_0 \otimes (\mbox{id},I)$ is irreducible.

There are two eigenspaces of $\hat{I}$: one to the eigenvalue $+1$, consisting
of the pairs with $\psi_u(p) = \psi_v(p)$, the other to the eigenvalue $-1$,
containing the pairs with $\psi_u(p) = -\psi_v(p)$. If we choose one of these
eigenspaces as our final Hilbert space, we obtain a representation of the
classical algebra on the quotient space. Again, the two possible choices give
equivalent theories. The final result can easily be brought to the following
form. The states are determined by complex functions $\varphi(p)$ on ${\mathbf
  R}_+$; the scalar product $(\varphi,\psi)$ is
\[
  (\varphi,\psi) = \int_0^\infty\frac{dp}{p}\ \varphi^*(p)\psi(p);
\]
let us denote the corresponding Hilbert space by $\mathcal K$. The
representatives of the above algebra are
\begin{eqnarray*}
 (\hat{p}_t\varphi)(p) & = & -p\varphi(p), \\
 (\hat{p}_r^2\varphi)(p) & = & p^2\varphi(p), \\
 (\hat{D}\varphi)(p) & = & -ip\frac{d\varphi(p)}{dp}, \\
 (\hat{J}^2\varphi)(p) & = & -p\frac{d\varphi(p)}{dp} -
 p^2\frac{d^2 \varphi(p)}{dp^2}. 
\end{eqnarray*}

The next question is that of time evolution. Time evolution of a generally
covariant system described by Dirac observables may seem self-contradictory or
gauge dependent. Here, we apply the approach that has been worked out in
\cite{honnef} and \cite{evol} using the symmetry group of time shifts found in
Sec.\ 2 of Ref.\ \cite{H-kief}, which is generated by the function $p_t$. The
operator $-\hat{p}_t$ has the meaning of the total energy $M$ of the system.
We observe that it is a self-adjoint operator with a positive spectrum and
that it is diagonal in our representation. The parameter $t$ of the unitary
group $\hat{U}(t)$ that is generated by $-\hat{p}_t$ is easy to interpret: $t$
represents the quantity that is conjugated to $p_t$ in the classical theory
and this is given by Eq.\ (\ref{tr}). Hence, $\hat{U}(t)$ describes the
evolution of the shell states between the levels of the function $(U+V)/2$ on
$\mathcal M$.

The missing piece of information of where the shell is on $\mathcal M$ is
carried by the quantity $r$ of Eq.\ (\ref{tr}). We try to define the
corresponding position operator in three steps.

First, we observe that $r$ itself is not a Dirac observable, but the boost $J$
is, and that the value of $J$ at the surface $t=0$ coincides with $rp_t$. It
follows that the meaning of the Dirac observable $Jp_t^{-1}$ is the position
at the time $t=0$. This is in a nice correspondence with the Newton-Wigner
construction on one hand, and with the so-called evolving constants of motion
by Rovelli \cite{rov2} on the other.

Second, we try to make $Jp_t^{-1}$ into a symmetric operator on our Hilbert
space. As it is odd with respect to $I$, we have to square it. Let us then
chose the following factor ordering:
\begin{equation}
 \hat{r}^2 := \frac{1}{\sqrt{p}}\hat{J}\frac{1}{p}\hat{J}\frac{1}{\sqrt{p}} =
 -\sqrt{p}\frac{d^2}{dp^2}\frac{1}{\sqrt{p}}.
\label{rsym}
\end{equation}
Other choices are possible; the above one makes $\hat{r}^2$
essentially a Laplacian and this simplifies the subsequent
mathematics. Indeed, we can map $\mathcal K$ unitarily to
$L^2({\mathbf R}_+)$ by sending each function $\psi(p) \in {\mathcal
K}$ to $\tilde{\psi}(p) \in L^2({\mathbf R}_+)$ as follows:
\[
  \tilde{\psi}(p) = \frac{1}{\sqrt{p}}\psi(p).
\]
Then, the operator of squared position $\tilde{r}^2$ on $L^2({\mathbf
R}_+)$ corresponding to $\hat{r}^2$ is
\[
  \tilde{r}^2 = \frac{1}{\sqrt{p}}\hat{r}^2\Bigl(\sqrt{p}\tilde{\psi}(p)\Bigr) =
  - \frac{d^2\tilde{\psi(p)}}{dp^2} = -\tilde{\Delta}\tilde{\psi}(p).
\]

Third, we have to extend the operator $\hat{r}^2$ to a self-adjoint
one. The Laplacian on the half-axis possesses a one-dimensional family
of such extensions \cite{reed-s}. The parameter is $\alpha \in
[0,\pi)$ and the domain of $\tilde{\Delta}_\alpha$ is defined by the
boundary condition at zero:
\[
  \tilde{\psi}(0) \sin\alpha + \tilde{\psi}'(0) \cos\alpha = 0.
\]
The complete system of normalized eigenfunctions of $\tilde{\Delta}_\alpha$
is given by:
\[
  \tilde{\psi}_\alpha(r,p) = \sqrt{\frac{2}{\pi}}\ \frac{r\cos\alpha\cos
  rp - \sin\alpha\sin rp}{\sqrt{r^2\cos^2\alpha + \sin^2\alpha}};
\]
if $\alpha \in (0,\pi/2)$, there is one additional bound state,
\[
  \tilde{\psi}_\alpha(b,p) = \frac{1}{\sqrt{2\tan\alpha}}\exp(-p\tan\alpha),
\]
so that
\begin{eqnarray*}
  -\tilde{\Delta}_\alpha\tilde{\psi}_\alpha(r,p) & = &
  r^2\tilde{\psi}_\alpha(r,p), \\
  -\tilde{\Delta}_\alpha\tilde{\psi}_\alpha(b,p) & = &
  -\tan^2\alpha\,\tilde{\psi}_\alpha(r,p).
\end{eqnarray*}
The corresponding eigenfunctions $\psi_\alpha$ of the operator
$\hat{r}^2_\alpha$ are:
\[
  \psi_\alpha(r,p) = \sqrt{\frac{2p}{\pi}}\ \frac{r\cos\alpha\cos
  rp - \sin\alpha\sin rp}{\sqrt{r^2\cos^2\alpha + \sin^2\alpha}},
\]
and we restrict ourselves to $\alpha \in [\pi/2,\pi]$, so that there
are no bound states and the operator $\hat{r}$ is self-adjoint.

To restrict the choice, we apply the idea of Newton and Wigner. First, the
subgroup of $G_0$ that preserves the surface $t = 0$ is to be found. This is,
in our case, $U_D(\lambda)$ generated by the dilatation $D$. Then, in the
quantum theory, the eigenfunctions of the position at $t=0$ are to transform
properly under this group; this means that the eigenfunction for the
eigenvalue $r$ is to be transformed to that for the eigenvalue
$U_D(\lambda)r$, for each $\lambda$.  The dilatation group generated by
$\hat{D}$ acts on a wave function $\psi(p)$ as follows:
\[
  \psi(p) \mapsto U_D(\lambda)\psi(p) = \psi(e^{-\lambda}p),
\]
where $U_D(\lambda)$ is an element of the group parameterized by
$\lambda$. Applying $U_D(\lambda)$ to $\psi_\alpha(r,p)$ yields
\[
  U_D(\lambda)\psi_\alpha(r,p) =
  e^{-\lambda/2}\sqrt{\frac{2p}{\pi}}\ \frac{r\cos\alpha\cos
  (e^{-\lambda}rp) - \sin\alpha\sin
  (e^{-\lambda}rp)}{\sqrt{r^2\cos^2\alpha + \sin^2\alpha}}.
\]
The factor $e^{-\lambda/2}$ in the resulting functions of $p$
keeps the system $\delta$-normalized.

Let $\alpha = \pi/2$; then
\[
  U_D(\lambda)\psi_{\pi/2}(r,p) = e^{-\lambda/2}\psi_{\pi/2}(e^{-\lambda}r,p).
\]
Similarly, for $\alpha = \pi$,
\[
  U_D(\lambda)\psi_{\pi}(r,p) = e^{-\lambda/2}\psi_{\pi}(e^{-\lambda}r,p),
\]
but such relation can hold for no other $\alpha$ from the interval
$[\pi/2,\pi]$, because of the form of the eigenfunction dependence on $r$. Now,
Newton and Wigner require that 
\[
  U_D(\lambda)\psi(r,p) = e^{-\lambda/2}\psi(e^{-\lambda}r,p).
\]  
Then all values of $\alpha$ except for $\alpha = \pi/2$ and $\alpha =
\pi$ are excluded.

We have, therefore, only two choices for the self-adjoint extension of
$\hat{r}^2$:
\begin{equation}
  \psi(r,p) := \sqrt{\frac{2p}{\pi}}\sin rp,\quad r \geq 0,
\label{rsa}
\end{equation}
and 
\[
  \psi(r,p) := \sqrt{\frac{2p}{\pi}}\cos rp,\quad r \geq 0.
\]
Let us select the first set, Eq.\ (\ref{rsa}); by that, the
construction of a position operator is finished.

The construction contains a lot of choice: the large factor-ordering
freedom, and the freedom of choosing the self-adjoint extension. One
can react to this ambiguity in two ways.

The first is to ask how the different choices influence the results. It seems
plausible that the qualitative, rough properties of the quantum system will be
the same for all possible choices. We hope (provisionally) that this is true.

The second question to ask is how the position is, in fact, measured
in praxis. This question hits the crux of the problem. Indeed, the
Newton-Wigner construction may be formally elegant but, to my
knowledge, nobody managed to describe the corresponding
measurement. If we search for methods of how the position of various
constituents in a microscopic system is measured, we find the
scattering method to dominate. For that it is necessary to use a
particular coupling the system under study, a crystal, say, has with
another agent, X-rays, say. One has to send the X-rays onto the
crystal and to view what comes out.

It seems, therefore, that the following approach would be more
reliable than attempts at a formal definition of a position operator
of the shell. One can try, for example, to couple the shell to some
field, the quanta of which could be emitted by the shell on its way
down and up. The quanta will, or will not reach the asymptotic
observers and their properties at infinity might reveal something of
what is going on with the shell. This is a future project because it
will be mathematically more difficult than our provisional attempt
with the position operator.

Another observable that we shall need is $\hat{\eta}$; this is to tell us the
direction of motion of the shell at the time zero, having the eigenvalues $+1$
for all purely outgoing shell states, and $-1$ for the in-going ones. In fact,
in the classical theory, $\eta = -\mathrm{sgn}p_r$, but $p_r$ does not act as
an operator on the Hilbert space $\mathcal K$, only $p_r^2$. Hence, we need
the following trick.

Consider the classical dilatation generator $D = tp_t + rp_r$. It is a
Dirac observable; at $t = 0$, its value is $rp_r$. Thus, for positive
$r$, the sign of $-D$ at $t = 0$ has the required value. On the
quotient space, the values at negative $r$ correspond to the
$I$-mapped states with positive $r$, and, as $D$ is $I$-invariant, the
relation of the sign to the direction of motion is again valid. Hence,
we have the relation:
\[
  \text{sgn}D = -\eta_{t=0}.
\]
 
The normalized eigenfunctions $\psi_a(p)$ of the operator $\hat{D}$
are solutions of the differential equation:
\[
  \hat{D}\psi_a(p) = a\psi_a(p).
\]
The corresponding normalized system is given by
\[
  \psi_a(p) = \frac{1}{\sqrt{2\pi}}p^{ia}.
\]
Hence, the kernels $P_\pm(p,p')$ of the projectors $\hat{P}_\pm$ on
the purely out- or in-going states are:
\[
  P_+(p,p') =
  \int_{-\infty}^0da\,\psi_a(p)\frac{\psi_a^*(p')}{p'},\quad P_-(p,p')
  = \int_0^\infty da\, \psi_a(p)\frac{\psi_a^*(p')}{p'}
\]
so that
\[
  (\hat{\eta}\psi)(p) = \int_0^\infty dp' [P_+(p,p') - P_-(p,p')]\psi(p').
\]

This finishes our construction of the shell quantum mechanics.

\section{Motion of wave packets}
We shall work with the family of wave packets on the energy half-axis that are
defined by
\[
  \psi_{\kappa\lambda}(p) := \frac{(2\lambda)^{\kappa+1/2}}{\sqrt{(2\kappa)!}}
  p^{\kappa+1/2}e^{-\lambda p},
\]
where $\kappa$ is a positive integer and $\lambda$ is a positive number with
dimension of length. Using the formula
\begin{equation}
  \int_0^\infty dp\,p^ne^{-\nu p} = \frac{n!}{\nu^{n+1}},
\label{22'1}
\end{equation}
which is valid for all non-negative integers $n$ and for all complex
$\nu$ that have a positive real part, we easily show that the wave
packets are normalized,
\[
  \int_0^\infty\frac{dp}{p}\ \psi_{\kappa\lambda}^2(p) = 1.
\]
The expected energy, 
\[
  \overline{M}_{\kappa\lambda} := \int_0^\infty\frac{dp}{p}\
  p\psi_{\kappa\lambda}^2(p),
\]
of the packet can be calculated by the same formula with the simple result
\[
 \overline{M}_{\kappa\lambda} = \frac{\kappa+1/2}{\lambda}. 
\]
The (energy) width of the packet can be represented by the mean
quadratic deviation, $\overline{\Delta M}_{\kappa\lambda}$, which is
\[
  \overline{\Delta M}_{\kappa\lambda} = \frac{\sqrt{2\kappa+1}}{2\lambda}.
\]
Hence, by choosing $\kappa$ and $\lambda$ suitably, we can approximate
any required energy and width arbitrarily closely.

The time evolution of the packet is generated by $-\hat{p}_t$:
\[
  \psi_{\kappa\lambda}(t,p) = \psi_{\kappa\lambda}(p) e^{-ipt}.
\]

Let us calculate the corresponding wave function
$\Psi_{\kappa\lambda}(r,t)$ in the $r$-representation,
\[
 \Psi_{\kappa\lambda}(t,r) := \int_0^\infty\frac{dp}{p}\ 
 \psi_{\kappa\lambda}(t,p)\psi(r,p),
\]
where the functions $\psi(r,p)$ are defined by Eq.\
(\ref{rsa}). Formula (\ref{22'1}) then yields:
\begin{equation}
  \Psi_{\kappa\lambda}(t,r) = \frac{1}{\sqrt{2\pi}}
  \frac{\kappa!(2\lambda)^{\kappa+1/2}}{\sqrt{(2\kappa)!}}
  \left[\frac{i}{(\lambda +it +ir)^{\kappa+1}} - \frac{i}{(\lambda +it
  -ir)^{\kappa+1}}\right]. 
\label{17:1}
\end{equation}
It follows immediately that
\[
  \lim_{r\rightarrow 0}|\Psi_{\kappa\lambda}(t,0)|^2 = 0.
\]
The scalar product measure for the $r$-re\-pre\-sentation is just $dr$ because
the eigenfunctions (\ref{rsa}) are normalized, so the probability to find the
shell between $r$ and $r+dr$ is $|\Psi_{\kappa\lambda}(t,r)|^2dr$.

Our first important result is, therefore, that the wave packets start
away from the center $r=0$ and then are keeping away from it during
the whole evolution. This can be interpreted as the {\em absence of
singularity} in the quantum theory: no part of the packet is squeezed
up to a point, unlike the shell in the classical theory.

Observe that the equation $\Psi_{\kappa\lambda}(t,0) = 0$ is {\em not} a
result of a boundary condition imposed on the wave function. It is a
result of the unitary dynamics. The nature of the question that we are
studying requires that the wave packets start in the asymptotic region so that
their wave function vanishes at $r = 0$ for $t \rightarrow -\infty$; this is
the only condition put in by hand. The fact that the dynamics preserves this
equation is the property of the unique self-adjoint extension of the
Hamiltonian operator.

A more tedious calculation is needed to obtain the time dependence
$\bar{r}_{\kappa\lambda}(t)$ of the expected radius of the shell,
\begin{equation}
  \bar{r}_{\kappa\lambda}(t) := \int_0^\infty
  dr\,r|\Psi_{\kappa\lambda}(t,r)|^2.
\label{17:2}
\end{equation}
Let first \underline{$\kappa = 0$}. The wave function of the packet
then is
\[
  \Psi_{0\lambda}(t,r) = 2\sqrt{\frac{\lambda}{\pi}}\ \frac{r}{r^2 +
  (\lambda + it)^2},
\]
so the expectation value of $\hat{r}$ is
\[
  \bar{r}_{0\lambda}(t) = \frac{4\lambda}{\pi}\int_0^\infty
  dr\,\frac{r^3}{(r^2 + \lambda^2 - t^2)^2 + 4\lambda^2 t^2}.
\]
This integral diverges logarithmically, so 
\[
  \bar{r}_{0\lambda}(t) = \infty.
\]

Let \underline{$\kappa \neq 0$}. The substitution of Eq.\ (\ref{17:1})
into (\ref{17:2}) leads to:
\[
  \bar{r}_{\kappa\lambda}(t) = \frac{1}{2\pi}
  \frac{(\kappa!)^2(2\lambda)^{2\kappa +
  1}}{(2\kappa)!}\Bigl(I_{\kappa\lambda}(t) -
  J_{\kappa\lambda}(t)\Bigr),
\]
where
\begin{eqnarray*}
 I_{\kappa\lambda}(t) & = & \int_0^\infty
  r\,dr\left\{\frac{1}{[(r+t)^2 + \lambda^2]^{\kappa+1}} +
  \frac{1}{[(r-t)^2 + \lambda^2]^{\kappa+1}}\right\}, \\
 J_{\kappa\lambda}(t) & = & \int_0^\infty
  r\,dr\left\{\frac{1}{[(\lambda-ir)^2 + t^2]^{\kappa+1}} +
  \frac{1}{[(\lambda+ir)^2 + t^2]^{\kappa+1}}\right\}.
\end{eqnarray*}
The first integral can be brought by elementary methods to the
following form:
\[
  I_{\kappa\lambda}(t) =
  \frac{1}{\kappa}\frac{1}{(t^2+\lambda^2)^\kappa} +
  t\int_{-t}^t\frac{ds}{(s^2+\lambda^2)^{\kappa+1}}.
\]

Let us calculate the second integral. We obtain after a simple
rearrangement:
\begin{eqnarray*}
  J_{\kappa\lambda}(t) & = & (-1)^{\kappa+1}\int_0^\infty
  dr\left\{\frac{r+i\lambda}{[(r+i\lambda)^2 - t^2]^{\kappa+1}} 
  -\frac{i\lambda}{[(r+i\lambda)^2 - t^2]^{\kappa+1}}\right. \\
  && \left. +\frac{r-i\lambda}{[(r-i\lambda)^2 - t^2]^{\kappa+1}} 
  +\frac{i\lambda}{[(r-i\lambda)^2 - t^2]^{\kappa+1}}\right\}.
\end{eqnarray*}
This suggests the introduction of integration contours $C_1$
defined in the complex plane by $z = r+i\lambda$ for $r\in
(0,\infty)$, and $C_2$ by $z = r-i\lambda$, $r\in (0,\infty)$. Then
$J_{\kappa\lambda}(t)$ can be written as follows:
\begin{eqnarray*}
  J_{\kappa\lambda}(t) & = & (-1)^{\kappa+1}\int_{C_1}
  dz\left[\frac{z}{(z^2 - t^2)^{\kappa+1}} -\frac{i\lambda}{(z^2 -
  t^2)^{\kappa+1}}\right] \\ &&
  (-1)^{\kappa+1}\int_{C_2}\left[\frac{z}{(z^2
  - t^2)^{\kappa+1}} +\frac{i\lambda}{(z^2 - t^2)^{\kappa+1}}\right].
\end{eqnarray*}
The integrals of the first terms in the square brackets can be done
immediately:
\begin{equation}
  J_{\kappa\lambda}(t) =
-\frac{1}{\kappa}\frac{1}{(\lambda^2+t^2)^\kappa} +
(-1)^{\kappa+1}i\lambda
\int_{-C_1+C_2}\frac{dz}{(z^2-t^2)^{\kappa+1}}.
\label{28'1}
\end{equation}

We obtain as the final result:
\begin{eqnarray}
  \lefteqn{\bar{r}_{\kappa\lambda}(t) =
  \frac{1}{2\pi}\frac{(\kappa!)^2(2\lambda)^{2\kappa+1}}{(2\kappa)!}
  \left[\frac{2}{\kappa}\frac{1}{(\lambda^2+t^2)^\kappa} +
  t\int_{-t}^t \frac{dx}{(x^2 + \lambda^2)^{\kappa+1}}\right.} \nn \\
  && \left. \mbox{} + i\lambda(-1)^{\kappa+1}\int_{C_1}\frac{dz}{(z^2 
  - t^2)^{\kappa+1}} - i\lambda(-1)^{\kappa+1}\int_{C_2}
  \frac{dz}{(z^2 - t^2)^{\kappa+1}}\right].
\label{30'1}
\end{eqnarray}
In fact, the R. H. side diverges for $\kappa = 0$ so, in this sense,
this formula can be considered as completely general, ie., valid for
all $\kappa$ and $t$.
 
Let us study some properties of the function
$\bar{r}_{\kappa\lambda}(t)$. Eq.\ (\ref{30'1}) implies that
\[
 \bar{r}_{\kappa\lambda}(t) = \bar{r}_{\kappa\lambda}(-t),
\]
so the average motion of the packet is symmetric under time
reversal. Eq.\ (\ref{30'1}) is also suitable for the calculation of
the expansions about the points $t =0$ and $t = \pm\infty$. Consider
first the point $t = 0$. Expanding the first term in the square
bracket is easy:
\[
  \frac{2}{\kappa}\frac{1}{(\lambda^2 + t^2)^\kappa} =
  \frac{2}{\kappa\lambda^{2\kappa}} \sum_{k=0}^\infty
  (-1)^k\left(\begin{array}{c} \kappa + k - 1 \\
  k\end{array}\right)\left(\frac{t}{\lambda}\right)^{2k}.
\] 
The series on the R. H. side converges for $|t| < \lambda$.

To expand the next term, we expand the integrand in the powers of
$x/\lambda$; the series converges for $|t| < \lambda$. Integrating
term by term yields:
\[
  t\int_{-t}^t \frac{dx}{(x^2 + \lambda^2)^{\kappa+1}} =
  \frac{2}{\lambda^{2\kappa}} \sum_{k=0}^\infty
  \frac{(-1)^k}{2k+1}\left(\begin{array}{c} \kappa + k \\
  k\end{array}\right)\left(\frac{t}{\lambda}\right)^{2k+2}.
\]
Again, this series converges for $|t| < \lambda$. 

A similar method can be applied to the remaining integrals:
\[
  \frac{1}{(z^2 - t^2)^{\kappa+1}} = \frac{1}{z^{2\kappa+2}}\sum_{k=0}^\infty
  \left(\begin{array}{c} \kappa + k \\
  k\end{array}\right)\left(\frac{t}{z}\right)^{2k}.
\]
The convergence is granted for $|t| < |z|$. As the minimal $|z|$ along
both contours is $\lambda$, the expansion is always valid for $|t| <
\lambda$. Then 
\[
  i\lambda(-1)^{\kappa+1}\int_{C_1 - C_2}\frac{dz}{(z^2 
  - t^2)^{\kappa+1}} = -\frac{2}{\lambda^{2\kappa}} \sum_{k=0}^\infty
  \frac{(-1)^k}{2\kappa+2k+1}\left(\begin{array}{c} \kappa + k \\
  k\end{array}\right)\left(\frac{t}{\lambda}\right)^{2k}.
\]
Collecting all terms, we obtain the expansion around $t=0$,
\begin{eqnarray}
  \lefteqn{\bar{r}_{\kappa\lambda}(t) =
  \frac{\lambda}{\pi}\frac{(\kappa!)^22^{2\kappa+1}}{(2\kappa)!}
  \left[\frac{\kappa+1}{\kappa(2\kappa+1)}+\right.}
  \nn \\
  &&\left. (\kappa+1) \sum_{k=1}^\infty
  \frac{(-1)^{k+1}}{(2\kappa+2k+1)k(2k-1)}\left(\begin{array}{c} \kappa + k - 1\\
  k\end{array}\right)\left(\frac{t}{\lambda}\right)^{2k}\right],
\label{28}
\end{eqnarray}
and the equation holds for $|t|<\lambda$. As the $k=1$ term in Eq.\
(\ref{28}) is positive, there is a minimal expected radius
$\bar{r}_{\kappa\lambda}(0)$ at $t=0$,
\begin{equation}
  \bar{r}_{\kappa\lambda}(0) =
  \frac{1}{\pi}\frac{2^{2\kappa}(\kappa!)^2}{(2\kappa)!}
  \frac{\kappa+1}{\kappa}\frac{\lambda}{\kappa+1/2} > 0.
\label{r0}
\end{equation}
For a large $\kappa$, the minimum at $t=0$ may be only local and/or
the curve may oscillate for $t\in (-\lambda,\lambda)$.

Let us turn to the asymptotics $t\rightarrow\pm\infty$. It is
sufficient to consider the case $t\rightarrow\infty$ because the other
one is obtained by $t\mapsto -t$. The first term in the square
brackets in Eq.\ (\ref{30'1}) is clearly of the order
$O(t^{-2\kappa})$ and it is not difficult to convince oneself that the
last two terms are both of the order $O(t^{-2\kappa-1})$.

The second term need more care. First, we use the relation
\[
  \int_{-t}^t \frac{dx}{(x^2+\lambda^2)^{\kappa+1}} =
  \frac{1}{\kappa!}\left(-\frac{1}{2\lambda}\frac{d}{d\lambda}\right)^\kappa
  \int_{-t}^t\frac{dx}{x^2+\lambda^2}
\]
so that we obtain
\[
  t\int_{-t}^t \frac{dx}{(x^2+\lambda^2)^{\kappa+1}} =
  \frac{2}{\kappa!}
  \left(-\frac{1}{2\lambda}\frac{d}{d\lambda}\right)^\kappa
  \left(\frac{t}{\lambda}\text{arctan}\frac{t}{\lambda}\right).
\]

For $t/\lambda>0$, the following formula holds:
\[
  \text{arctan}\frac{t}{\lambda} = \frac{\pi}{2} -
  \text{arctan}\frac{\lambda}{t}.
\]
Using it and expanding the function arctan$(\lambda/t)$ around zero leads to
\[
  t\int_{-t}^t \frac{dx}{(x^2+\lambda^2)^{\kappa+1}} =
  \frac{\pi t}{2^\kappa\kappa!}
  \left(-\frac{1}{\lambda}\frac{d}{d\lambda}\right)^\kappa\frac{1}{\lambda}
  - 2(-1)^\kappa\left(\frac{d}{d(\lambda)^2}\right)^\kappa
  \sum_{k=0}^\infty\frac{(-1)^k}{2k+1}\left(\frac{\lambda}{t}\right)^{2k}.
\]
The series converges for $t>\lambda$ and can be differentiated term by
term in this interval. The first non zero term comes only from
$k=\kappa$ and it has the value
\[
  -\frac{2\kappa!}{2\kappa+1}\frac{1}{t^{2\kappa}}.
\]
It holds also
\[
  \left(-\frac{1}{\lambda}\frac{d}{d\lambda}\right)^\kappa\frac{1}{\lambda} = 
  \frac{(2\kappa-1)!!}{\lambda^{2\kappa+1}}.
\]
Hence,
\[
  t\int_{-t}^t \frac{dx}{(x^2+\lambda^2)^{\kappa+1}} = \pi
  t\frac{(2\kappa-1)!!}{2^\kappa\lambda^{2\kappa+1}\kappa!} +
  O(t^{-2\kappa}).
\]
Substituting this into Eq.\ (\ref{30'1}) and using the symmetry $t\mapsto
-t$, we obtain for both cases $t\rightarrow \pm\infty$:
\begin{equation}
  \bar{r}_{\kappa\lambda}(t) \approx |t| + O(t^{-2\kappa}).
\label{rinfty}
\end{equation}

A further interesting question about the motion of the packets is
about the portion of a given packet that moves in---is purely
in-going---at a given time $t$. The portion is given by
$\|\hat{P}_-\psi_{\kappa\lambda}\|^2$, where $\hat{P}_-$ is the
projector defined in Sec.\ \ref{sec:quant}. Let us calculate this quantity.

If we write out the projector kernel and make some simple
rearrangements in the expression of the norm, we obtain:
\[
  \|\hat{P}_-\psi_{\kappa\lambda}\|^2 = \int_{-\infty}^\infty dq'
  \int_{-\infty}^\infty dq'' \left(\int_0^\infty
  da\,\psi^*_a(e^{q'})\psi_a(e^{q''})\right)
  \psi^*_{\kappa\lambda}(t,e^{q''}) \psi_{\kappa\lambda}(t,e^{q'}),
\]
where the transformation of integration variables $p'$ and $p''$ to $e^{q'}$
and $e^{q''}$ in the projector kernels has been performed.

The integral in the parenthesis,
\[
  \int_0^\infty da\,\psi^*_a(e^{q'})\psi_a(e^{q''}) = \frac{1}{2\pi}
  \int_0^\infty da\, e^{ia(q''-q')},
\]
is a kernel in an integral that is exponentially damped at the
infinities. Thus, we can calculate it as a limit,
\begin{eqnarray*}
  \frac{1}{2\pi}\lim_{\epsilon\rightarrow 0}\int_0^\infty da\,
  e^{ia(q''-q')-\epsilon a} & = &
  \frac{i}{2\pi}\lim_{\epsilon\rightarrow 0} \frac{1}{(q''-q')
  +i\epsilon} \\ & = & \frac{i}{2\pi}{\mathcal P}\frac{1}{q''-q'} +
  \frac{1}{2}\delta(q''-q'),
\end{eqnarray*}
where $\mathcal P$ denotes the principal value.

Doing the integral over the $\delta$-function gives the simple result:
\[
  \frac{1}{2}(\psi_{\kappa\lambda},\psi_{\kappa\lambda}) = \frac{1}{2}.
\]
The rest can be written as follows:
\begin{eqnarray*}
  \lefteqn{\|\hat{P}_-\psi_{\kappa\lambda}\|^2 = \frac{1}{2} +} \\
  && \frac{i}{2\pi}\int_{-\infty}^\infty dq' {\mathcal
  P}\int_{-\infty}^\infty dq''
  \psi_{\kappa\lambda}(e^{q'})\psi_{\kappa\lambda}(e^{q''}) \frac{\cos
  t(q''-q') + i\sin t(q''-q')}{q''-q'}.
\end{eqnarray*}
The integrand is a sum of a symmetric and an anti-symmetric functions
of the variables $q'$ and $q''$. The principal value integral
annihilates the anti-symmetric part. The integral from the symmetric
part is already regular, and we can write the final formula:
\begin{equation}
  \|\hat{P}_-\psi_{\kappa\lambda}\|^2 = \frac{1}{2} -
  \frac{1}{2\pi}\int_{-\infty}^\infty dq' \int_{-\infty}^\infty dq''
  \psi_{\kappa\lambda}(e^{q'})\psi_{\kappa\lambda}(e^{q''}) \frac{\sin
  t(q''-q')}{q''-q'}.
\label{3:1}
\end{equation}

Let us calculate the in-going portion for some simple values of
$t$. Thus, for $t=0$, we obtain immediately:
\[
  \|\hat{P}_-\psi_{\kappa\lambda}\|^2_{t=0} = \frac{1}{2}.
\]
At the time zero, the probabilities to catch the shell going in or out
are equal.

The limit $t\rightarrow\pm\infty$ can be obtained, if we use the formula:
\[
  \lim_{t\rightarrow\pm\infty}\frac{\sin tx}{x} = \pm\pi\delta(x).
\]
Hence,
\[
  \lim_{t\rightarrow\pm\infty}\frac{\sin t(e^{q''}-e^{q'})}{q''-q'} =
  \pm\pi\frac{e^{q''}-e^{q'}}{q''-q'}\delta(e^{q''}-e^{q'}).
\]
Substituting this into the integral of Eq.\ (\ref{3:1}) and returning
back to the variables $p'$ and $p''$ results in:
\[
  \|\hat{P}_-\psi_{\kappa\lambda}\|^2_{t\rightarrow\pm\infty} =
  \frac{1}{2} \mp \frac{1}{2} \int_0^\infty\frac{dp'}{p'}
  \int_0^\infty\frac{dp''}{p''} \psi_{\kappa\lambda}(p')
  \psi_{\kappa\lambda}(p'') \frac{p''-p'}{\log p'' - \log p'}
  \delta(p''-p').
\]
The expression
\[
  \frac{p''-p'}{\log p'' - \log p'}
\]
is smooth and equal to  $p'$ at $p''=p'$. Hence, finally
\[
  \|\hat{P}_-\psi_{\kappa\lambda}\|^2_{t\rightarrow -\infty} = 1,\quad
  \|\hat{P}_-\psi_{\kappa\lambda}\|^2_{t\rightarrow \infty} = 0,
\]
and we have only in-going, or only outgoing shells at the infinity.

The obvious interpretation of these formulae is that quantum shell always
bounces at the center and re-expands. We can, however, ask further questions.
For example, what is the time delay of the re-expansion as compared, say, with
the same trajectory in the background manifold $\mathcal M$ that carries the
flat metric
\begin{equation}
 ds^2 = -dUdV + (1/4)(-U+V)^2 d\Omega^2
\label{fm}
\end{equation}
in our coordinates $U$ and $V$? To find this time delay, the ``true'' metric
with respect to these coordinates had to be calculated. The metric is
determined by the quantum state in a similar way as the position and the
colour of the horizon are (see the next section). However, unlike the points
and the metric inside $\mathcal M$, the points and metric in the asymptotic
region are gauge invariant quantities. The method by which we should calculate
the asymptotic metric ought to make the gauge invariance of the result
transparent. Such a method has first to be developed.

The result that the quantum shell bounces and re-expands is clearly at variance
with the classical idea of black hole forming in the collapse and preventing
anything that falls into it from re-emerging. It is, therefore, natural to ask,
if the packet is squeezed enough so that an important part of it comes under
its Schwarzschild radius. We can try to answer this question by comparing the
minimal expected radius $\bar{r}_{\kappa\lambda}(0)$ with the expected
Schwarzschild radius $\bar{r}_{\kappa\lambda H}$ of the wave packet. The
Schwarzschild radius is given by
\[
 \bar{r}_{\kappa\lambda H} = 2\mbox{G}\bar{M}_{\kappa\lambda} =
 2\frac{\bar{M}_{\kappa\lambda}}{M_P^2},
\]
where $M_P$ is the Planck energy. Now, the values of $\kappa$ and $\lambda$
for which a large part of the packet gets under its Schwarzschild radius
clearly satisfy the inequality
\[
  \bar{r}_{\kappa\lambda}(0) < \bar{r}_{\kappa\lambda H},
\]
or
\begin{equation}
  (\lambda M_P)^2 < 2\pi\frac{\kappa(\kappa + 1/2)^2}{\kappa + 1}
  \frac{(2\kappa)!}{2^{2\kappa}(\kappa!)^2}.
\label{squee1}
\end{equation}

Interpreting $\lambda$ roughly as the spatial width of the packet, we
have $\lambda M_P \gg 1$ for reasonably broad packets. Then the
right-hand side can be estimated by the Stirling formula:
\[
 2\pi\frac{\kappa(\kappa + 1/2)^2}{\kappa + 1}
  \frac{(2\kappa)!}{2^{2\kappa}(\kappa!)^2} \approx \sqrt{2\pi}\kappa.
\]
Substituting this into the inequality (\ref{squee1}) yields
\begin{equation}
  \bar{M}_{\kappa\lambda} > \frac{\lambda M_P}{\sqrt{2\pi}}M_P,
\label{squee2}
\end{equation}
which implies that the threshold energy for squeezing the packet under its
Schwarz\-schild radius is much larger than the Planck energy. For narrow wave
packets, we have that $\lambda M_P \approx 1$, so the inequality
(\ref{squee1}) is satisfied, and the threshold energy is about one Planck
energy. The inequality (\ref{squee2}) expresses, therefore, always the desired
property. To summarize: Reasonably narrow packets can, in principle, get under
their Schwarzschild radius; their energy must be much larger than Planck
energy.  Even in such a case, the shell bounces and re-expands.

This apparent paradox will be explained in the next section.

\section{Grey horizons}
In this section, we try to explain the apparently contradictory result that
the quantum shell can cross its Schwarzschild radius in both directions. The
first possible idea that comes to mind is simply to disregard everything
that our model says about Planck regime. This may be justified, because the
model can hardly be considered as adequate for this regime.  However, the
model {\em is} mathematically consistent, simple and solvable; it must,
therefore, provide some mechanism to make the horizon leaky. We shall study
this mechanism in the hope that it can work in more realistic situations, too.

To begin with, we have to recall that the Schwarzschild radius is the radius
of a non-diverging null hyper-surface; anything moving to the future can cross
such a hyper-surface only in one direction. The local geometry is that of an
apparent horizon. (Whether or not an event horizon forms, that can also depend
on the geometry near the singularity \cite{H-contra}). However, as Einstein's
equations are invariant under time reversal, there are two types of
Schwarzschild radius: that associated with a black hole and that associated
with a white hole. Let us call these Schwarzschild radii themselves {\em
  black} and {\em white}. The explanation of the paradox that follows from the
model is that quantum states can contain a linear combination of black and
white horizons, and that no event horizon forms. We call such a combination a
{\em grey horizon}.

The existence of grey horizons can be shown as follows. The position and the
``colour'' of a Schwarzschild radius outside the shell is determined by the
spacetime metric. For our model, this metric is a combination of purely gauge
and purely dependent degrees of freedom, and so it is determined, within the
classical version of the theory, by the physical degrees of freedom through
the constraints.

To explain the idea in more detail, less us start with the general case in the
ADM formalism. There are 16 canonical variables, the 6 components of the
three-metric $q_{kl}$, the 6 components of the conjugate momentum $\pi^{kl}$,
1 lapse and three shift functions. These can be decomposed (non-uniquely) into
physical, gauge, and dependent variables. Fixing the gauge variables by hand
(this also includes some boundary conditions in non-compact cases) means that
a particular space-like surface $\Sigma$ is chosen, and a particular
coordinate system $x^k$, $k= 1,2,3$, is lain onto this surface. Then the
constraints turn into differential equations determining the dependent part of
$q_{kl}$ and $\pi^{kl}$ in terms of the physical one and so the tensor fields
$q_{kl}$ and $\pi^{kl}$ are determined uniquely along $\Sigma$ in the
coordinates $x^k$ by the physical degrees of freedom. By this, the full
spacetime metric $g_{\mu\nu}$ and all its first derivatives are known at each
point of $\Sigma$. Indeed, if we choose the Gaussian coordinates $x^0$, $x^k$,
adapted to $\Sigma$, then the four-metric at $\Sigma$ is
\[
  ds^2 = -(dx^0)^2 + q_{kl}dx^kdx^l,
\]
and the derivatives of this metric with respect to the coordinates
$x^0$ and $x^k$ are given by
\begin{alignat}{3}
  \frac{\partial g_{00}}{\partial x^0} & = 0, & \qquad \frac{\partial
  g_{00}}{\partial x^k} & =  0, & \qquad \frac{\partial
  g_{0k}}{\partial x^0} & =  0, \nn \\ \frac{\partial g_{kl}}{\partial
  x^0} & = -2K_{kl}, & \qquad \frac{\partial g_{kl}}{\partial x^m} & =
  \frac{\partial q_{kl}}{\partial x^m}, & \qquad \frac{\partial
  g_{0k}}{\partial x^l} & =  0, \nn
\end{alignat}
where $K_{kl} := (\det q_{kl})^{-1/2}(1/2\,q^{mn}\pi_{mn}
q_{kl}-\pi_{kl})$ is the second fundamental form of the surface
$\Sigma$. Observe that the choice of Gaussian coordinates is
equivalent to specifying the lapse and shift at $\Sigma$ by hand. The
lapse and shift could also be fixed by the condition that the gauge is
preserved by the evolution.

Let $S$ be a closed two-surface on $\Sigma$. We can calculate the Gaussian
coordinates adapted to $S$ in $\Sigma$ from the metric $q_{kl}$ of $\Sigma$;
let they be $x^{\prime A}$ and $x^{\prime 3}$, $A = 1,2$, so that $S$ is given
by $x^{\prime 3} = 0$ and $x^{\prime 3}$ increases in the outside direction.
Let the corresponding components of the tensor fields be $q'_{kl}$ and
$K'_{kl}$. Then the induced two-metric on $S$ is $q'_{AB}$ and the full
three-metric on $\Sigma$ is
\[
  ds^2 = (dx^{\prime 3})^2 + q'_{AB}dx^{\prime A}dx^{\prime B}.
\]
Now, let $l^\mu$ and $n^\mu$ be null vectors orthogonal to $S$ ,
$l^\mu$ being the outgoing and $n^\mu$ the in-going one. Their
component in terms of the coordinates $x^{\prime 0} := x^0$ and
$x^{\prime k}$ are
\[
  l^{\prime\mu} = (1,0,0,1),\quad n^{\prime\mu} = (1,0,0,-1).
\]
Then
\[
  \frac{\partial g'_{AB}}{\partial x^{\prime\mu}}l^{\prime\mu} =
  \frac{\partial g'_{AB}}{\partial x^{\prime 0}} + \frac{\partial
  g'_{AB}}{\partial x^{\prime 3}} = -2K'_{AB} + \frac{\partial
  q'_{AB}}{\partial x^{\prime 3}},
\]
and, similarly,
\[
  \frac{\partial g'_{AB}}{\partial x^{\prime\mu}}l^{\prime\mu} =
  -2K'_{AB} - \frac{\partial q'_{AB}}{\partial x^{\prime 3}},
\]
We can, therefore, check, whether or not the following equation holds 
\begin{equation}
  g^{\prime AB}\left(-2K'_{AB} \pm \frac{\partial
  q'_{AB}}{\partial x^{\prime 3}}\right) = 0,
\label{3''1}
\end{equation}
and so can find, if $S$ is an out- (in-)going apparent horizon---Eq.\
(\ref{3''1}) is then valid with the above (lower) sign.

Let us look to see how this algorithm works for the shell model. The
constraints are $P_U(\rho) = 0$ and $P_V(\rho)=0$. If the
transformation from the original variables $\Lambda(\rho)$, $R(\rho)$,
$P_\lambda(\rho)$, $P_R(\rho)$, $\eta$, $\mathbf r$ and $\mathbf p$ to
$U(\rho)$, $V(\rho)$, $P_U(\rho)$ $P_V(\rho)$, $\eta$, $u$, $v$, $p_u$
and $p_v$ were known, it would provide the functionals:
\[
  P_U(\rho) = P_U[\lambda,R,P_\lambda,P_R,\eta,{\mathbf r},{\mathbf p};\rho),
\]
and
\[
  P_V(\rho) = P_V[\lambda,R,P_\lambda,P_R,\eta,{\mathbf r},{\mathbf p};\rho).
\]
The transformation is not known explicitly, but we know that the
constraint equations
\begin{equation}
  P_U[\lambda,R,P_\lambda,P_R,\eta,{\mathbf r},{\mathbf p};\rho) =
  0,\quad P_V[\lambda,R,P_\lambda,P_R,\eta,{\mathbf r},{\mathbf
  p};\rho) = 0
\label{4''1}
\end{equation}
are equivalent to the original constraints, Eqs.\ (\ref{LWF-H}) and
(\ref{LWF-Hr}). Hence, our first trick is to work with Eqs.\
(\ref{LWF-H}) and (\ref{LWF-Hr}) instead of Eqs.\ (\ref{4''1}).

The constraints (\ref{LWF-H}) and (\ref{LWF-Hr}) contain the physical
variables $\eta$, $M$ $u$ and $v$ also through ${\mathbf r}$ and $\mathbf
p$. We can choose the gauge variables to be $R(\rho)$ and
$\Lambda(\rho)$. A fixed function $R(\rho)$ determines $\rho$ in terms
of the geometrical quantity $R$ and so it fixes a radial coordinate along
$\Sigma$. $\Lambda(\rho)$ contains derivatives of the embedding
functions, so it determines the slope of the embedding at each
$\rho$. Integrating the slope gives a family of surfaces; a
suitable boundary condition at infinity selects one of them.

In order to obtain a suitable surface $\Sigma$ the functions $R(\rho)$
and $\Lambda(\rho)$ have to satisfy some further boundary conditions at the
infinity, at the shell and at the regular center. The condition at the
infinity, $\rho \rightarrow \infty$, is to guarantee that $\Sigma$ is
asymptotically flat. That at the shell is necessary in order that
$\Sigma$ is smooth across the shell. Finally, at $\rho = 0$, we
require that $\Sigma$ cut the regular center rather that the
singularity and that it be a smooth surface at this point. Similarly,
$P_R$ and $P_\Lambda$ have to satisfy suitable boundary conditions at
$\rho = 0$, $\rho = {\mathbf r}$ and $\rho \rightarrow \infty$. The
explicit form of these boundary conditions are carefully discussed in
\cite{L-W-F}.

Then the constraints become equations for the two functions $P_R(\rho)$
and $P_\Lambda(\rho)$. Eq.\ (\ref{LWF-H}) is an algebraic equation for
$P_R$; solving it and inserting the solution into Eq.\ (\ref{LWF-Hr})
gives an ordinary differential equation for $P_\Lambda$. The
differential equation, together with the boundary conditions,
determines $P_\Lambda(\rho)$ uniquely, and this, in turn, together
with Eq.\ (\ref{LWF-H}), gives $P_R(\rho)$. The solution is unique. From the
known functions $R(\rho)$, $\Lambda(\rho)$, $P_R(\rho)$ and $P_\Lambda(\rho)$,
we can determine $q_{kl}$ and $K_{kl}$ along $\Sigma$ and check Eq.\
(\ref{3''1}). 

One can try to solve Eqs.\ (\ref{LWF-H}) and (\ref{LWF-Hr}) for $P_R(\rho)$
and $P_\Lambda(\rho)$ explicitly, by choosing the functions $R(\rho)$ and
$\Lambda(\rho)$ in some way that simplifies the equations. Instead, we use the
uniqueness of the solution in the following simple trick. Any solution of the
constraint equations {\em in the spherically symmetric case} defines an
initial data and surface for a solution to Einstein's equations that is itself
spherically symmetric. Hence, every such solution of constraints forms a
space-like surface that can be embedded in some Schwarzschild spacetime. There
will always be the Schwarzschild solution of mass zero inside the shell, and
the Schwarzschild solution of mass $M$ outside it.

In this way, we find by inspection from the Kruskal diagram: If the shell is
in-going, $\eta = -1$, then it is contracting and any space-like surface
containing such a shell can at most intersect an {\em outgoing} apparent
horizon at the radius $R = 2\text{G}M$, independently of which of the two
infinities the surface is connecting the shell with. Analogous result holds
for $\eta=+1$, where the shell is expanding. The corresponding $\rho_H$ is
determined by the equation $R(\rho_H) = 2\text{G}M$, and the horizon will cut
$\Sigma$ if and only if $\rho_H > {\mathbf r}$. We can assign the value $+1$
($-1$) to the horizon that is out- (in-)going and denote the quantity by $c$
(colour: black or white hole). Then $c = -\eta$.

In particular, if we choose the gauge so that $R(\rho) = \rho$, then
$\mathbf r$ is just $r(t)$, where $t$ is the value of the parameter
$t$ at which the shell intersect $\Sigma$, and we have: 
\begin{enumerate}
\item The condition that an apparent horizon intersects $\Sigma$ is
  $r_t < 2\text{G}M$.
\item The position of the horizon at $\Sigma$ is $\rho_H = 2\text{G}M$.
\item The value of $c$ is $c = -\eta$.
\end{enumerate}
 
In this way, questions about the existence and colour of an
apparent horizon outside the shell are reduced to equations containing
dynamical variables of the shell. In particular, the result that
$c=-\eta$ can be expressed by saying that the shell always creates a
horizon outside that cannot block its motion. All that matters is that
the shell can bounce at the singularity (which it cannot within the
classical theory).

These results can be carried over to quantum mechanics after quantities such
as $2\mbox{G}M - r$ and $\eta$ are expressed in terms of the operators
describing the shell. Then we obtain a ``quantum horizon'' with the ``expected
radius'' $2\mbox{G}\bar{M}$ and with the ``expected colour'' $-\bar{\eta}$ to
be mostly black at the time when the expected radius of the shell crosses the
horizons inwards, neutrally grey at the time of the bounce and mostly white
when the shell crosses it outwards.

This proof has, however, two weak points. First, the spacetime metric on the
background manifold is not a gauge invariant quantity; although all gauge
invariant geometrical properties can be extracted from it within the classical
version of the theory, this does not seem to be possible in the quantum theory
\cite{haj2}. Second, calculating the quantum spacetime geometry along
hyper-surfaces of a foliation on a given background manifold is foliation
dependent. For example, one can easily imagine two hyper-surfaces $\Sigma$ and
$\Sigma'$ belonging to different foliations, that intersect each other at a
sphere outside the shell and such that $\Sigma$ intersects the shell in its
in-going and $\Sigma'$ in its outgoing state. Observe that the need for a
foliation is only due to our insistence on calculating the quantum metric.

The essence of these problems is the gauge dependence of the results
of the calculation. However, it seems that this dependence concerns only
details such as the distribution of different hues of grey along the horizon,
not the qualitative fact that the horizon exists and changes colour from
almost black to almost white. Still, a more reliable method to establish the
existence and properties of grey horizons would require another material
system to be coupled to our model; this could probe the spacetime geometry
around the shell in a gauge-invariant way.

It may still seem difficult to imagine any spacetime that contains an
apparent horizon of mixed colours. Nevertheless, examples of such
space-times can readily be constructed if the assumption of
differentiability is abandoned. A continuous, piecewise differentiable
spacetime can make sense as a history within the path integral method.

The simplest construction of this kind is based on the existence of the time
reversal isometry $\mathcal T$ as defined in the foregoing paper \cite{H-kief}
that maps an in-going shell spacetime onto an outgoing one.

Let us choose a space-like hyper-surface $\Sigma_1$ crossing the shell
before this hits the singularity in a $(-1,M,u)$-spacetime $\mathcal
M$, and find the corresponding surface ${\mathcal T}\Sigma_1$ in the
spacetime $\mathcal TM$ with the parameters $(1,M,v)$.  Then we cut
away the part of $\mathcal M$ that lies in the future of $\Sigma_1$
and the part of $\mathcal TM$ in the past of ${\mathcal
T}\Sigma_1$. As their boundaries are isometric to each other, the
remaining halfs can be stuck together in a continuous way. In the
resulting spacetime, the shell contracts from the infinity until it
reaches $\Sigma_1$ at the radius $r_1$; then, it turns its motion
abruptly to expand towards infinity again. There is no singularity and
the spacetime is flat everywhere inside the shell. If $r_1 <
2\mbox{G}M$, then there is an apparent horizon at $R = 2\mbox{G}M$. It
comes into being where the in-going shell crosses the radius $r =
2\mbox{G}M$ and is outgoing (black) until it reaches $\Sigma_1$.
Then, it changes its colour abruptly to white (in-going) and lasts only
until the outgoing shell crosses it again.

The space-like hyper-surface $\Sigma_1$ can be chosen arbitrarily in
$\mathcal M$. The construction can, therefore, be repeated in the
future of ${\mathcal T}\Sigma_1$ in an analogous way so that we obtain
a spacetime with two ``pleats''; the shell contracts, then expands,
then contracts again and hits the singularity. The horizon starts as a
black ring, then changes to a white one, and then it becomes black for
all times. This history is, however, not continuous. Clearly, one can
repeat the construction arbitrary many times; this leads to a
``pleated'' spacetime with a zig-zag motion of the shell and
alternating horizon rings of white and black colour. If the spacetime
is to be singularity free, however, there must be an odd number of
pleats and an even number of rings, beginning with the black ring and
ending with a white one.

The conditions that the surface $\Sigma_1$ cuts the trajectory of the shell at
some small value of the Schwarzschild radial coordinate $R$, is smooth and
space-like everywhere and hits the space-like infinity $i^0$ for large values
of $R$ allow a considerable freedom. We can require in addition that
$\Sigma_1$ joins smoothly to the surface $T = T_1$, where $T$ is the
Schwarzschild time coordinate and $T_1$ some constant so that $\Sigma_1$
coincides with $T = T_1$ for all values of $R$ larger than, say, $R_1$. It is
clear from the Penrose diagram that such a $\Sigma_1$ can ran arbitrarily
close to the incoming shell trajectory and can be joined to $T = T_1$ for
arbitrary low value of $T_1 \in (-\infty,\infty)$, if $R_1$ is chosen
sufficiently large. On the other hand, for $R_1 = 2\text{G}M + \epsilon$,
$\Sigma_1$ can join $T = T_1$ for arbitrarily large $T_1 \in
(-\infty,\infty)$, just if $\epsilon > 0$.

Consider now an observer at the fixed value $R_0$ of the Schwarzschild radius
in each shell spacetime. We shall choose $\Sigma_1$ in such a way that $R_1 <
R_0$. With this choice, the observer trajectory $R = R_0$ remains smooth at
$\Sigma_1$. Then, the lower bound on the possible values of $T_1$ is $T_c$,
which is the Schwarzschild time of the point at which the observer crosses the
shell. There is, however, no upper bound on $T_1$. Hence, we can construct a
one-pleat spacetime for each value of $T_1$ from the interval $(T_c,\infty)$
with a smooth trajectory of the observer. For each value of $T_1$, the
observer will measure the proper time
\[
  \Delta\tau = 2\sqrt{\left(1-\frac{2\text{G}M}{R_0}\right)}(T_1 - T_c) \in
  (0,\infty)
\]
between his two encounters with the shell. Thus, the time delay can be made
arbitrarily small or large. (Of course, all such histories and many others
must be integrated with some suitable measure in a path integral in order to
obtain a reasonable value of the delay).

Let us choose a gauge in each spacetime constructed above such that the
trajectory of the shell is $V = u$ for the in-going part and $U = u$ for the
outgoing one. Then, the metric in the asymptotic region, where the observer
is, will read
\[
  ds^2 = -A(U,V)dUdV + R^2(U,V)dR^2,
\]
and it is clear that the functions $A(U,V)$ and $R(U,V)$ must have different
forms for different values of $T_1$, or else the proper time $\Delta\tau$
measured by the observer will be independent of $T_1$. In most cases, the
asymptotic behaviour of the metric in this gauge will be different from
(\ref{fm}). On the other hand, in each such spacetime, there will be double
null coordinates $U_1$ and $V_1$, say, in which the metric will have the
asymptotic behaviour (\ref{fm}). However, the trajectory of the shell will
then be given, with respect to the coordinates $U_1$ and $V_1$, by different
equations for different values of $T_1$.

\section{Concluding remarks}
Comparison of the motion of wave packets of Sec.\ 4 with the classical
dynamics of the shell as described in Sec.\ 3 of \cite{H-kief} shows a marked
difference. Whereas all classical shells cross their Schwarzschild radius and
reach the singularity in some stage of their evolution, the quantum wave
packets never reach the singularity, but always bounce and re-expand; few of
them manage to cross their Schwarzschild radius during their motion. This
behaviour is far from being a small perturbation around a classical solution
if the classical spacetime is considered as a whole. Even locally, the
semi-classical approximation is not valid near the bouncing point.  It is
surely valid in the whole asymptotic region, where narrow wave packets follow
more or less the classical trajectories of the shell.

The most important question, however, concerns the validity of the
semi-classical approximation near the Schwarzschild radius. We have seen that
the geometry near the radius can resemble the classical black hole geometry in
the neighbourhood of the point where the shell is crossing the Schwarzschild
radius inwards. Then, the radius changes its colour gradually and the geometry
becomes very different from the classical one. Finally, near the point where
the shell crosses the Schwarzschild radius outwards, the radius is
predominantly white and the quantum geometry can be again similar to the
classical geometry, this time of a white hole horizon.

If the change of colour is very slow then the neighbourhood of the inward
crossing where the classical geometry is a good approximation can be large. It
seems that sufficiently large time delays would allow for arbitrarily slow
change of colour. We cannot exclude, therefore, that the quantum spacetime
contains an extended region with the geometry resembling its classical
counterpart near a black hole horizon, at least locally. This can be true even
if the quantum spacetime as a whole differs strongly from any typical
classical collapse solution.

One can even imagine the following scenario (which needs a more realistic
model than a single thin shell). A quantum system with a large energy
collapses and re-expands with huge time delay. The black hole horizon phase is
so long, that Hawking evaporation becomes significant and must be taken into
account in the calculation. It does then influence the time delay and the
period of validity of the black hole approximation. The black hole becomes
very small and only then the change of horizon colour becomes significant. The
white hole stage is quite short and it is only the small remnant of the system
that, finally, re-expands. The whole process can still preserve unitarity. In
fact, this is a scenario for the issue of Hawking evaporation process. At
least, it is not excluded by the results of the present paper.

The calculations of this paper are valid only for null shells. Similar
calculations have been performed in \cite{H-K-K}. There has been re-expansion
and unitarity for massive shells if the rest mass has been smaller than the
Planck mass ($10^-5$ g). It is very plausible that the interpretation of these
results is similar to that given in the present paper. Thus, we can expect the
results valid at least for all ``light'' shells. There is, in any case, a long
way to any astrophysically significant system and a lot of work is to be done
before we can claim some understanding of the collapse problem.

Our method of dealing with the problem employs simplified models and a kind
of effective theory of gravity; it does not worry about the final form of a
full-fledged theory of quantum gravity. This need not be completely
unreasonable approach. Even if the ultimate quantum gravity theory were known,
most calculations would still be performed within the approximation of some
effective theory and for simplified models (compare the situation in the QCD).
The method can give useful hints also because of the fact that the black hole
geometry is ``made up'' from purely dependent degrees of freedom of the
gravitational field, and these degrees of freedom have no proper quantum
character of their own.

To summarize: We have demonstrated, at least for light shells, that quantum
theory can smoothly unify two states of motion, one being the time reversal of
the other, into one history. In this way, geometry containing a piece of a
black hole horizon can be followed by geometry containing a piece of a white
hole horizon---just the opposite to the situation we know from the Kruskal
diagram of the classical general relativity. In this way, the quantum
evolution can stay unitary and the question posed at the beginning of the paper
can be answered as follows: A quantum system is not {\em always} lost if it
falls under its Schwarzschild radius.

\subsection*{Acknowledgments}
Part of this work has been done at the University of Utah, where the author
enjoyed a nice hospitality. Helpful discussions with K.~V.~Kucha\v{r},
C.~Kiefer and V.~F.~Mukhanov are acknowledged. The work was supported by the
Swiss National Fonds, the Tomalla Foundation Z\"{u}rich, and the NSF Grant
PHY-9734871 to the University of Utah.

\end{document}